\documentclass[aps,prl,twocolumn,groupedaddress]{revtex4}
\usepackage{graphicx}
\usepackage{bm}
\usepackage{bbm}
\usepackage{epic}
\usepackage{eepic}
\usepackage{pifont}
\usepackage{nicefrac}
\usepackage{amsmath}
\usepackage{amssymb}
\begin{document}

\title{Absence of low temperature anomaly on the melting curve
of $^4$He}

\author{I.\,A.\,Todoshchenko}

\author{H.\,Alles} 

\author{H.\,J.\,Junes} 

\author{A.\,Ya.\,Parshin}

\altaffiliation{P.\,L.\,Kapitza Institute, Kosygina 2, Moscow
119334, Russia}

\author{V.\,Tsepelin}

\altaffiliation{Department of Physics, Lancaster University,
Lancaster, LA1 4YB, UK}

\affiliation{Low Temperature Laboratory, Helsinki University of
Technology, P.\,O.\,Box 2200, FIN-02015~HUT, Finland}

\email{todo@boojum.hut.fi}

\date{\today}

\begin{abstract}
We have measured the melting pressure and pressure 
in the liquid at constant density of ultra-pure $^4$He 
(0.3\,ppb of $^3$He impurities) with the accuracy of about 
0.5\,$\mu$bar 
in the temperature range from 10 to 320\,mK. Our measurements 
show that the anomaly on the melting curve below 80\,mK which 
we have recently observed \cite{T06} is entirely due to an 
anomaly in the elastic modulus of Be-Cu from which our pressure 
gauge is made of. We thus conclude that the melting pressure of 
$^4$He follows the $T^4$ law due to phonons in the whole 
temperature range from 10 to 320\,mK without any sign of a 
supersolid transition.
\end{abstract}


\maketitle

Recent experimental results obtained by Kim and Chan 
\cite{KC04,KC05} have revived great interest to the problem of 
supersolidity which was first discussed
almost 40 years ago \cite{AL69,Ch70,Le70}. The supersolid state 
of matter is characterized by the coexistence of crystalline 
order and superfluidity. In helium crystals, according to Andreev 
and Lifshitz \cite{AL69} and Chester \cite{Ch70}, quantum 
delocalization of point defects (most probably -- vacancies) might 
decrease their activation energy to zero. Bose condensation of 
such defects can lead then to superfluidity in a crystal, 
that is, supersolidity. During 1970s and 1980s many experimental 
groups tried to detect this possible supersolid state by various methods, but unsuccessfully (see \cite{Me92} for a review).

One possible manifestation of supersolidity would be so called
nonclassical rotational inertia (NCRI) -- a reduction
in the rotational inertia of a solid at low temperatures 
\cite{Le70}, which can be detected by the torsional oscillator 
measurements on a helium crystal, similarly to the famous 
Andronikashvili experiment \cite{And46}. 
In 1981, Bishop, Paalanen and Reppy \cite{BPR81}
have carried out such measurements at temperatures down to 
25\,mK with carefully annealed helium samples. With
$\sim5\times10^{-6}$ sensitivity to the superfluid fraction 
$\rho_s/\rho$ they found no evidence of NCRI. However, recently 
Kim and Chan (KC) have
observed the effect of NCRI below 0.2\,K at different pressures,
from melting pressure up to 140\,bar, with $\rho_s/\rho\sim0.01$ 
in the low temperature limit. They interpreted the onset of NCRI 
at 0.2\,K as an indication of the transition of solid helium to 
the supersolid phase \cite{KC06}.

To date, several experimental groups have confirmed the KC observations \cite{RR06,PYK06,KTSS06}. Rittner and Reppy 
\cite{RR06} have also found that $\rho_s/\rho$ is not a 
universal characteristic of solid helium but can be 
reduced below a detectable level through annealing of the 
sample. Quite recently, they even have been able to produce 
highly disordered helium samples, where $\rho_s/\rho$ reached 
much higher magnitudes, up to 0.2 \cite{RR07}. However, other 
groups \cite{PYK06,KTSS06,KC06} have not confirmed the 
annealing effect.

In the absence of a consistent explanation of all available data 
on the annealing effect, the interpretation of torsional 
oscillator measurements in terms of a superflow in a $^4$He 
crystal remains controversial. Note in this connection that the 
observations of KC could be explained, at least in principle, by 
classical mechanism of dislocation-induced plasticity of solid 
helium. With appropriate temperature dependence of the dislocation 
mobility both a reduction in the rotational inertia  and a 
temperature peak in the damping of oscillations can be obtained, 
as in the Granato-L\"ucke theory \cite{GL56}. Similar 
interpretations have been suggested in \cite{BGNT06,NBGT06}.

At the moment, no other evidence for superflow in bulk solid
helium has been found. Recent searches for a pressure driven
superflow have given null results \cite{DHB05,DB06}. On the other
hand, Sasaki {\it et al.}\ \cite{SICMB06} have detected a
superflow presumably at grain boundaries in polycrystalline solid
helium in contact with the superfluid phase. They observed this
phenomenon even at temperatures as high as 1.1\,K, which is too 
high compared to the supposed supersolid transition temperature of
0.2\,K (and is not very far from the superfluid transition 
temperature in the bulk liquid). Thus it is not clear yet 
whether this interesting observation is relevant to the KC 
experiments.

If the observed onset of NCRI really is a manifestation of a phase
transition in the bulk solid, the equilibrium thermodynamic 
properties of the solid should also display an anomaly. The
magnitude of this anomaly in the case of Bose condensation of
vacancies can be estimated as $\delta S\sim R\rho_s/\rho$, 
where $\delta S$ is the excess entropy due to vacancies just 
above the transition temperature $T_c$ and $R$ is the gas 
constant. Below $T_c$ the excess entropy should drop to zero. 
With $\rho_s/\rho\sim10^{-3}\div10^{-2}$ it gives very large 
$\delta S$, which certainly has been ruled out by heat capacity
measurements \cite{Me92}. However, this naive estimate is 
valid only in the case of weakly interacting Bose gas. As an 
alternative, Anderson {\it et al.}\ \cite{ABH05} have suggested 
a model of the supersolid ground state with a number of strongly correlated vacancies and interstitials. In this model, there 
are no soft modes (in contrast to the Andreev-Lifshitz 
scenario), and $\delta S$ may be very small even with 
relatively large superfluid fraction. Another model of a ground 
state with a low density of strongly correlated 
vacancies/interstitials was suggested by Dai
{\it et al.}\ \cite{DMZ05}.

Thus at present there is no theoretical consensus on possible
magnitude of the entropy change associated with the supposed
supersolid transition. As for the experimental data, a few of
earlier measurements have revealed deviations from conventional 
Debye behavior below 0.5\,K, but later experiments have not 
reproduced any of such anomalies \cite{Me92}. The most recent 
studies \cite{CC06} have found a small excess heat capacity 
at temperatures down to 80\,mK, which corresponds to 
$\delta S\sim10^{-7}R$ at 300\,mK, however, no indication of 
a phase transition near 200\,mK has been found. The annealing 
effect has not been studied, and the nature of the observed 
anomaly has remained unclear. Thus it seems very important to 
look for any anomaly in the entropy of high-quality $^4$He 
crystals around 0.2\,K.

Recently we have reported on our direct high-precision
measurements of the melting pressure of $^4$He of regular purity
($\approx80$\,ppb of $^3$He impurities) in the range from 10 to 
400\,mK \cite{T06}. The melting pressure showed the expected 
$T^4$ dependence, and the coefficient was in excellent agreement 
with available data on the sound velocity in the liquid and the 
Debye temperature of the solid $^4$He. However, we have observed 
an anomaly below about 100\,mK, where $T^4$ dependence changed to 
much weaker, almost linear dependence. It was not clear at that 
time, what was the origin of the low temperature anomaly: 
influence of $^3$He impurities, effects due to change in
the crystal shape, or some other reason.

In this Letter we present a new set of high-precision 
measurements on the melting curve of $^4$He with ultra-pure 
($0.3\pm0.08$\,ppb of $^3$He impurities) 
\cite{KOK}
helium where we have used 
an interferometer \cite{JLTP} to monitor the crystal shape.
In addition, we have used a cryogenic valve to close the cell
containing only liquid $^4$He and measured the temperature
dependence of the sensitivity of the pressure gauge. We have 
found out that the low temperature anomaly which we have 
observed in our previous experiments with $^4$He of regular 
purity \cite{T06} is also present for high purity $^4$He sample, 
and that it is not due to the change in the crystal shape. 
However, the same anomaly was observed for the pressure which 
was measured in the liquid at a constant volume. Thus our 
results prove that the reason for the observed anomaly is the 
small ($2\times10^{-7}$) decrease of the spring constant of our 
pressure gauge membrane below 100\,mK. After applying the 
correction on the observed temperature dependence of the 
sensitivity of the gauge, the measured variation of the melting pressure of $^4$He below 320\,mK does not deviate from the pure 
phonon $T^4$ law with the accuracy of about 0.5\,$\mu$bar. This 
sets the upper limit of $\sim5\times10^{-8}R$ for a possible 
excess entropy in the solid $^4$He below 320\,mK.

Our capacitive pressure gauge, of a standard Straty-Adams design
\cite{Straty}, is made of beryllium bronze (Be-Cu) and has the
sensitivity $dC/dp=41$\,pF$/$bar at the melting pressure
of 25.31\,bar. The capacitance of the gauge was measured by the
commercial Andeen-Hagerling 2500A bridge yielding the accuracy of
about 0.5\,$\mu$bar after $\sim1$\,minute averaging. Temperature
was measured by a $^3$He melting curve thermometer anchored to the
sample cell. The $^3$He melting pressure was converted to
temperature according to the Provisional Low Temperature
Scale, PLTS-2000 \cite{PLTS2000}.

Crystals were nucleated and grown at constant temperature by 
slowly increasing the pressure in the cell. In order to create 
crystals in the 
field of view, we have used a capacitive nucleator which was 
operated with high voltage. During measurements on the melting pressure at different temperatures, crystals were imaged with 
a low-temperature Fabry-P$\rm\acute e$rot interferometer 
\cite{JLTP}. Crystals were typically grown to 
about 1\,cm in diameter, and during cooling or warming their 
height and curvature did not change remarkably, ensuring that no 
hydrostatic or capillary corrections to the measured pressure 
were needed.

\begin{figure}[t]
\centering
\includegraphics[width=1.0\linewidth]{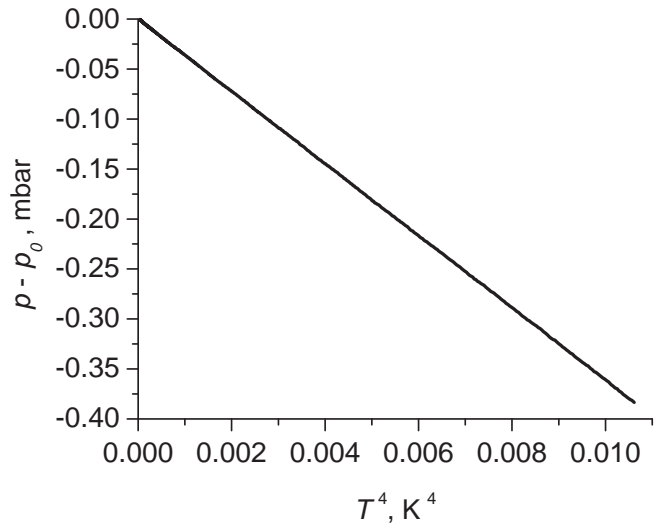}
\caption{\label{fig:t4}
Melting pressure of ultra-pure $^4$He below 320\,mK.
The presented curve consists of 1000 raw data points.
}
\end{figure}

\begin{figure}[t]
\centering
\includegraphics[width=1.0\linewidth]{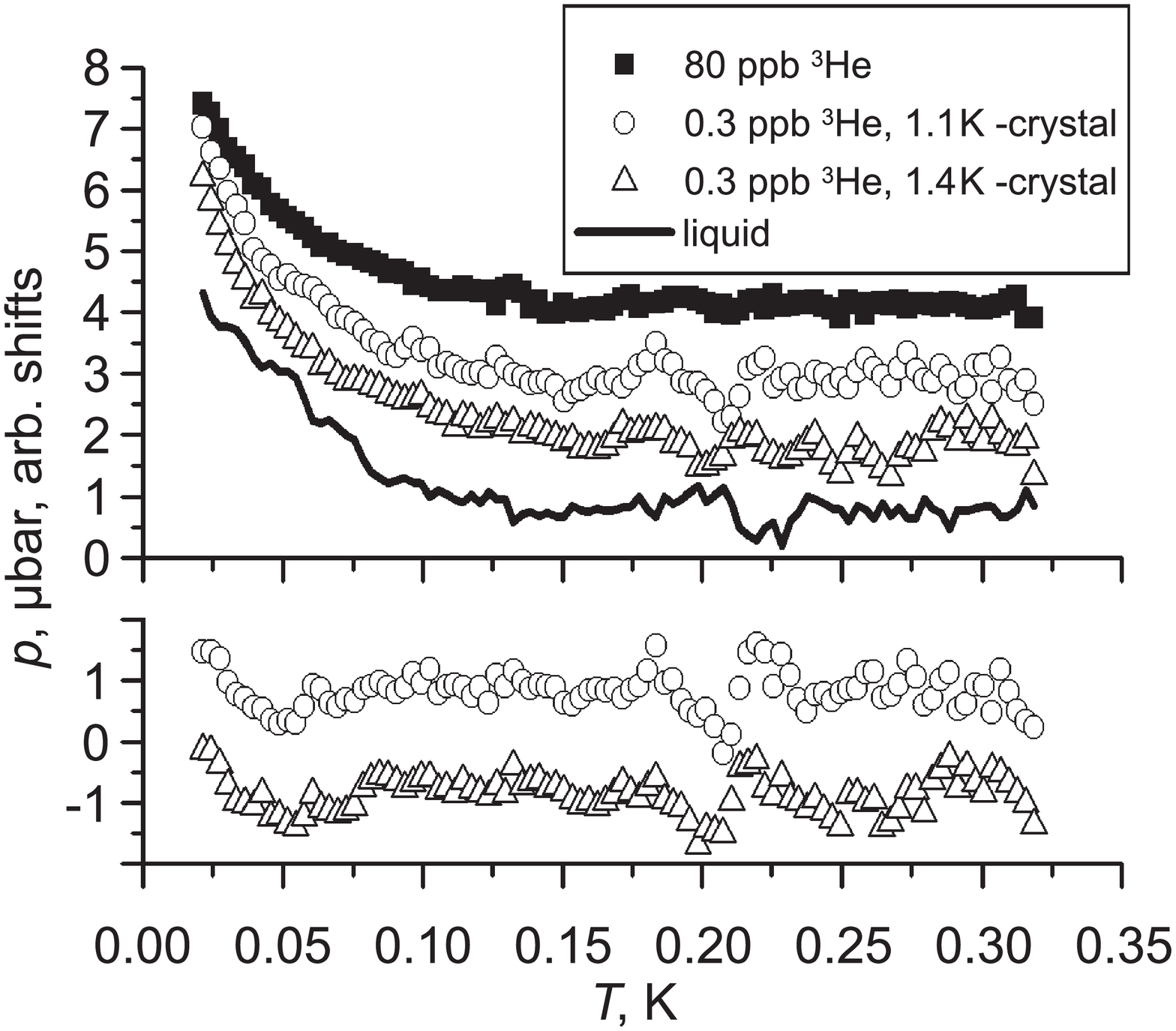}
\caption{\label{fig:substr}
{\it Upper part.}
Deviation of the measured pressures from the corresponding
$T^4$ fits:
melting pressure of $^4$He of regular purity ($\blacksquare$),
melting pressure of ultra-pure $^4$He measured with
crystal grown at 1.1\,K 
(\raise 0.04cm\hbox{\scriptsize $\bigcirc$}) and with  
crystal grown at 1.4\,K ($\triangle$), pressure in the liquid 
at a constant volume (solid line). 
{\it Lower part.} Deviation of the difference between the melting 
pressure and the pressure in the liquid at a constant volume 
from the best $T^4$ fit. Symbols refer to the same 
crystals as in the upper part. All curves are offset for clarity.
%In each curve the vertical shift is arbitrary. 
%Note that the subtracted $T^4$ terms vary by about 1\,mbar in 
%this temperature range.
}
\end{figure}

The results of the measurements on the melting pressure with
ultra-pure $^4$He were
similar to the results obtained with $^4$He of regular purity
\cite{T06}, see Figs.\,\ref{fig:t4} and \ref{fig:substr}. 
The sample crystals were nucleated and grown at 10\,mK, at 
1.1\,K where melting pressure is 80\,mbar higher than at low temperatures, and at 1.4\,K where melting pressure is 700\,mbar 
higher than at low temperatures. The
sensitivity of our pressure gauge is high enough to check 
(1) the effect of non-hydrostatic stresses \cite{TB92} produced 
by the change of the melting pressure when the crystals grown at 
1.4\,K were cooled down and (2) the effect of disorder due to 
possible plastic deformation of such crystals. However, after subtracting off the
best $T^4$ fit to the data, no other reproducible contribution to
the melting pressure is seen above 100\,mK with the accuracy of
$\sim0.5\,\mu$bar. It means that non-hydrostatic stresses in
samples grown at 1.4\,K had been effectively annealed during the
first cooldowns of such samples. The low temperature anomaly below
100\,mK is present for all ultra-pure and normal purity samples 
which proves that it is not due to $^3$He impurities.

Another possible reason for the anomaly could be the corresponding
temperature dependence of the pressure gauge sensitivity. The
flexible membrane of the pressure gauge is made of beryllium
bronze, widely used material for low temperature experiments
because of its high tensile strengh and low losses on mechanical
deformations. However, it is known that some types of beryllium
bronze may have low temperature anomaly in the heat capacity
\cite{Cryogenics} and Young modulus \cite{P02,Parpia}. The
anomaly we have observed in the melting pressure of $^4$He 
below 100\,mK would correspond to a very small, 
$\sim2\times10^{-7}$, relative decrease of the Young modulus of the 
bronze (which is still several orders of magnitude higher compared 
to conventional metals). The only way to detect so small change in 
the pressure gauge sensitivity is to measure the temperature
dependence of the capacitance of the gauge at a constant pressure, 
or, at a pressure which depends on temperature in a known way.

We have measured the pressure in liquid $^4$He just below the
melting curve at a constant volume. Our cryogenic valve, placed on the mixing chamber plate, was found to keep the filling line of the
cell closed with very small ($<0.5\,\mu$bar$/$day) leakage, thus
making careful measurements of the pressure in the liquid possible. The variation of the pressure at a constant volume is due to
thermal expansion of the liquid and can be expressed as

\begin{align*}
\left(\frac{\partial p_{_L}}{\partial T}\right)_V=-\frac{\rho}{V}
\left( \frac{\partial S}{\partial\rho}\right)_T.
\end{align*}

\noindent
In the low temperature limit the thermodynamics of the liquid is
dominated by phonons, $S\propto T^3/c^3$, which gives
$\left({\partial p_{_L}}/{\partial T}\right)_V=3uS/V$, where
$u=(\rho/c)(\partial c/\partial\rho)_T$ is the Gr\"{u}neisen
constant. As a result, the pressure of the liquid at a constant 
volume varies as $T^4$.

Indeed, in the range of 100\,...\,320\,mK the measured pressure in
the liquid obeys $T^4$ law, while below 100\,mK we again 
observed the same low temperature anomaly (see
Fig.\,\ref{fig:substr}), apparently due to the temperature-dependent
sensitivity of our pressure gauge. The measured capacitance $C$ of
the gauge can be written in the form $1/C=1/C_0-Ap/Y$, where $Y$ 
is the Young modulus of the bronze and $A$ is a constant which
depends on the gauge design. Thus a relative change $\delta Y/Y$
in the Young modulus results in the equal relative change 
$-\delta p/p$ in the measured pressure.

Below 100\,mK, where $T^4$ contribution to the pressure due
to thermal expansion of the liquid 
is only about 0.5\,$\mu$bar, the change of the sensitivity by
about 5\,$\mu$bar/25\,${\rm bar}=2\times10^{-7}$ is obvious. However,
there might be some small variation of the sensitivity also above
100\,mK which would be mixed with the real pressure change. The
correct way to eliminate the contribution of the pressure
gauge to the measured pressure is to subtract the data measured in
the liquid from the data measured on the melting curve and look
for a non-phonon contribution in the residual. The results of the
subtractions are shown in Fig.\,\ref{fig:substr} (lower part). With 
the accuracy of 0.5\,$\mu$bar, the measured melting pressure
of $^4$He can be described by the $T^4$ dependence due to phonons 
and no sign of any phase transition is seen.

To summarize, we have carried out high-precision measurements on 
the melting pressure of ultra-pure $^4$He down to 10\,mK with
several samples grown at different pressures. The shape and 
height of the crystals were carefully controlled by the
interferometric imaging so that no corrections for the Laplace
pressure and hydrostatic pressure were needed. All samples were
single crystals, without any signs of grain boundaries or other
macroscopic defects. The growth thresholds for crystals were less 
than 1\,$\mu$bar, which guarantees low density of dislocations 
(less than $10^2$\,cm$^{-2}$). We have calibrated the sensitivity 
of our pressure gauge with the accuracy of $2\times10^{-8}$ by 
measuring the variation of the pressure in the liquid $^4$He at a 
constant volume. This calibration allowed us to eliminate the low 
temperature anomaly observed below 100\,mK \cite{T06}. As a 
result, we have found that the melting pressure of $^4$He does 
not deviate from the $T^4$ law due to phonons with the accuracy 
of 0.5\,$\mu$bar and there is no sign of a supersolid transition
down to 10\,mK. 

In conclusion, we would like to stress that our measurements
do not rule out the possibility of a supersolid transition
in high-quality $^4$He crystals. They only set the upper limit 
of $\sim5\times10^{-8}R$ for the non-phonon entropy in such 
crystals at the melting pressure below 320\,mK.

\begin{acknowledgments}

We are grateful to M.\,Paalanen, J.\,M.\,Parpia and G.\,Volovik
for fruitful discussions. This work was supported by the EC-funded
ULTI project, Transnational Access in Programme FP6 (contract
\#RITA-CT-2003-505313) and by the Academy of Finland (Finnish
Centre of Excellence Programmes 2000--2005, 2006--2011, and the
Visitors Programme 112401).

\end{acknowledgments}

\bibliography{todoshchenko}

%\begin{thebibliography}{99}
%\bibitem{KOK} We are 
%grateful to K.\,O. Keshishev for providing us with a sample of 
%this ultra-pure helium. The analysis of the sample purity was 
%made by the US Bureau of Mines
%\end{thebibliography}

\end{document}